\documentstyle[12pt,psfig,a4]{article}

\newcommand{\be}{\begin{equation}}
\newcommand{\ee}{\end{equation}}
\newcommand{\bea}{\begin{eqnarray}}
\newcommand{\eea}{\end{eqnarray}}
\newcommand{\Mth}{M(atrix)-theory }
\newcommand{\MS}{M\"obius strip }
\newcommand{\Tr}{\mbox{Tr}}
\newcommand{\nn}{\nonumber}
\newcommand{\inv}[1]{{\cal I}_{#1}}
\newcommand{\Sh}[1]{{\cal S}_{#1}}
\newcommand{\fr}[2]{\frac{#1}{#2}}
\newcommand{\gym}{g_{YM}}
\newcommand{\bmat}{\left(\begin{array}}
\newcommand{\emat}{\end{array}\right)}

\newcommand{\newsection}[1]{
   \vspace{5mm}
   \pagebreak[3]
   \addtocounter{section}{1}
   \setcounter{equation}{0}
   \setcounter{subsection}{0}
   \setcounter{footnote}{0}
   \addcontentsline{toc}{section}{\protect\numberline{\arabic{section}}{#1}}
   \begin{center}    
   {\bf {\large \thesection. #1}}
   \end{center}
   \nopagebreak[4]
   \nopagebreak[4]}

\newcommand{\acknowledgements}{
   \vspace{15mm}
   \pagebreak[3]
   {\bf {\large Acknowledgements}}
   \nopagebreak
   \medskip
   \nopagebreak} 

\def\NPB#1#2#3{{\it Nucl.\ Phys.}\/ {\bf B#1} (19#2) #3}
\def\PLB#1#2#3{{\it Phys.\ Lett.}\/ {\bf B#1} (19#2) #3}
\def\PRD#1#2#3{{\it Phys.\ Rev.}\/ {\bf D#1} (19#2) #3}
\def\PRL#1#2#3{{\it Phys.\ Rev.\ Lett.}\/ {\bf #1} (19#2) #3}

\begin{document}

\addtolength{\baselineskip}{.5mm}  
\renewcommand{\theequation}{\thesection.\arabic{equation}}
\renewcommand{\thefootnote}{\fnsymbol{footnote}}
 
\begin{flushright}
October 1997\\
THU-97/27\\
hep-th/9710057\\
\end{flushright}
\vspace{2cm}
\thispagestyle{empty}
\begin{center}
{\large{Matrix Theory on Non-Orientable Surfaces}}\\[18mm]
{\bf Gysbert Zwart\footnote{e-mail address: zwart@fys.ruu.nl} }\\[7.5mm]
{\it Institute for Theoretical Physics\\ University of Utrecht\\ 
Princetonplein 5 \\ 3584 CC  Utrecht\\ The Netherlands}\\[18mm]
 
{\bf Abstract}
\end{center}
We construct the Matrix theory descriptions of M-theory on the M\"obius 
strip and the Klein bottle. In a limit, these provide the matrix string 
theories for the CHL string and an orbifold of type IIA string theory.\\
 
\vfill
 
\pagebreak

\newsection{Introduction}

Different string theories have been claimed to be related via one 
unifying theory, M-theory. This theory is supposed to reproduce the 
various weak coupling string theories in certain limits of its moduli space, 
and to give the correct interpolation in between, at finite values of the 
coupling (see e.g. \cite{Mth}). The 
proposal of matrix theory as a definition of M-theory in the infinite 
momentum frame \cite{matrix, B} has allowed many of the claims to be 
verified. The matrix formulations of type II and heterotic strings in ten 
dimensions were constructed in \cite{strings}, and 
compactifications to lower dimensions on tori and orbifolds 
were investigated as well (see \cite{B} and references therein).  

Here we would like to concentrate on compactifications of M-theory to nine 
dimensions on non-orientable surfaces: the M\"obius strip and the Klein 
bottle. Type II compactifications on these manifolds, in the form of 
certain orientifold models, were considered in \cite{DP,P}. These 
authors also argued which nine dimensional string theories would 
appear as limits of M-theory on these surfaces: the M\"obius strip yields 
the nine-dimensional CHL string \cite{CHL}, with gauge group 
$E_8$ (arising as the twisted sector living on the (single) boundary of 
the strip, following \cite{HW}), and the Klein bottle represents, in a 
suitable limit, a type 
IIA string in nine dimensions modded out by half a shift over the circle 
accompanied by the operation $(-1)^{F_L}$.   

We use the orientifold models to obtain the matrix description of 
these theories. The models are constructed as torus 
compactifications of 
matrix theory, modded out by an appropriate symmetry group consisting of 
a reflection and shifts over half a period of the circles. We first 
consider the M\"obius strip. We construct the $2+1$ dimensional gauge 
theory describing the dynamics of zero-branes on the M\"obius strip. The 
base manifold of the gauge theory is itself again a M\"obius strip. The 
construction is similar to that of heterotic matrix theory \cite{Het}. We 
show how in the limit of weak string coupling we indeed recover a string 
theory 
with chiral fermions producing one $E_8$ gauge group, the CHL string. 

Then we turn to the Klein bottle compactification. In this case the 
T-duality that we have to perform to construct the 
$2+1$ dimensional theory describing the dynamics is less straightforward, 
we have to use the original construction of \cite{T,GRT} to find the gauge 
theory. 
The base manifold of the gauge theory is 
{\em not} a Klein bottle. Rather, the geometric structure of the 
Klein bottle is reflected in the structure of the gauge fields; 
different modes of the fields turn out to satisfy different 
conditions. Finally we again find the type IIA string theory, 
modulo 
the required symmetry, as a limit of this three-dimensional model.  

After the preprint of this paper was made public on the archives, 
\cite{KR} appeared, which also studies the matrix descriptions of 
M-theory on the M\"obius strip and the Klein bottle. These authors find 
similar results to ours in the case of the M\"obius strip. For 
the Klein bottle, however, they argue that the gauge theory base space is a 
Klein bottle, and not a cylinder as we found. A possible 
resolution to this contradiction was suggested in \cite{HWW}. There it 
was found, using non-commutative geometry techniques, that the topology 
of the base space depends on the value of the background anti-symmetric 
tensor field $B$ through the original Klein bottle. In the absence of 
this background the result exactly coincides with our conclusions, i.e. the 
gauge theory base space is a cylinder with the same field content as 
presented in 
this paper. If a half-integral $B$-field Wilson surface is switched on, 
however, the result is a Klein bottle. Presumably in the work of 
\cite{KR} there is such a $B$-field, although it is not obvious to us 
where it enters their argument.  

\newsection{M-theory on a M\"obius strip}

In constructing the \Mth  compactified on a \MS  we will start from a 
related type IA theory studied by \cite{DP,P}, and consider the 
dynamics of D0-branes in that model. 

The IA theory in question is given by the type IIA theory compactified 
on a two torus, with radii $R_{1,2}$, divided out by the symmetries 
$\Omega\inv1$ and $\Sh1\Sh2$. Here $\Omega$ is the world sheet 
orientation reversal, $\inv1$ inverts the first coordinate, making the 
first circle into a line segment, and the $\Sh{i}$ are shifts in the 
compact directions by half a period:
$$
X_1\to X_1+\pi R_1,\quad X_2\to X_2+\pi R_2. 
$$
In the M-theory point of view $\Omega\inv1$ will become the inversion 
of the eleventh direction, and the resulting compactification manifold 
is the M\"obius strip. 

M-theory on a circle divided by the inversion of the circle was 
demonstrated to be equivalent to heterotic $E_8\times E_8$ string theory 
\cite{HW}. One $E_8$ factor lives on each ten-dimensional boundary. Weak 
coupling corresponds to shrinking the compact direction. In the present 
case we divide out one more symmetry, which exchanges the two boundaries 
($\Sh1$) and rotates by half a period in an extra compact direction 
($\Sh2$). In the heterotic string, this is exactly the operation 
producing a nine-dimensional CHL string \cite{CHL}, so we expect our 
matrix model to produce matrix CHL string theory in the limit of 
vanishing $R_1$. 

We will first review Dabholkar and Park's analysis of the 
orientifold model underlying the matrix description. On 
closed string states with momentum numbers $n_1,n_2$, the action of 
$\Sh1\Sh2$ is simply $(-1)^{n_1+n_2}$.  
The untwisted closed string spectrum is therefore that of the usual 
type IA theory in eight dimensions, except that those states with   
odd $n_1+n_2$ are projected out (note that this does not affect the 
massless spectrum). In addition there are (massive) twisted states, 
having half integer winding numbers. 

The open string spectrum is 
obtained by calculating the Klein bottle contribution to the 
RR-tadpoles. The amplitude with $\Omega\inv1$ in the trace gives the 
usual $32$ D8-branes, aligned along the $2$-direction. Their $X_1$ 
coordinates have to be compatible with both $\inv1$ and $\Sh1$; the 
maximal symmetry is $SO(16)$, obtained when sixteen D8-branes lie on 
the orientifold plane $X_1=0$ (and the other sixteen, by $\Sh1$ 
symmetry, on $X_1=\pi R_1$). At strong coupling the symmetry is 
expected to become $E_8$, similarly as in the regular IA theory. The 
$\Omega\inv1\Sh1\Sh2$ in the trace 
gives a contribution that vanishes in the long tube limit. Finally, 
the twisted channels vanish altogether, since they necessarily have 
non-vanishing winding number in the $2$-direction, incompatible with 
$\Omega$ in the trace. The resulting massless spectrum is the same as 
that of the CHL model 
in eight dimensions, and the two models were in fact claimed to be dual
by \cite{DP,P}. 

\begin{figure}
\centerline{\psfig{figure=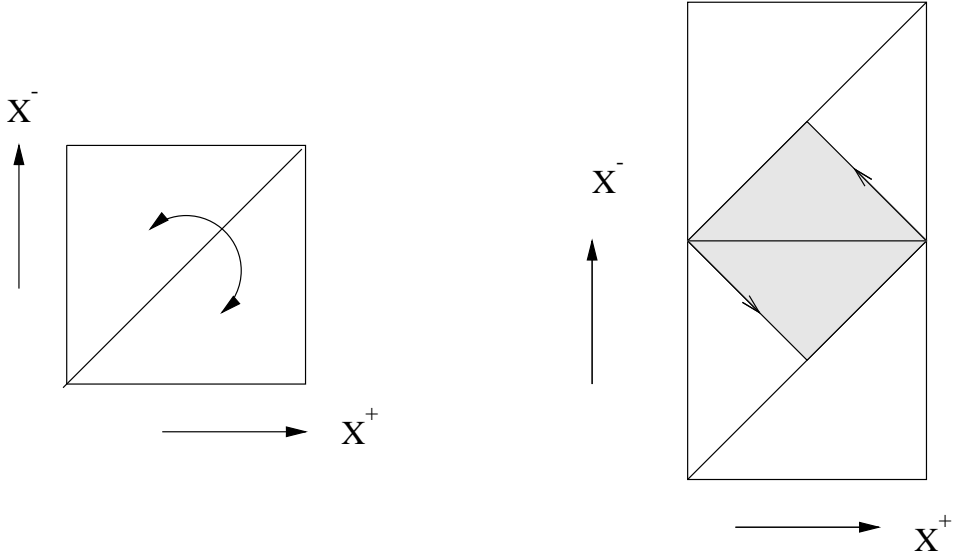}}
\caption{\it On the left, the torus is drawn, with the symmetry to be 
divided 
out. The result is a M\"obius strip. This can be most easily seen in the 
right-hand figure, where instead of the triangle we take the shaded 
region as fundamental domain; the arrows indicate the 
identification of the two sides} 
\label{fig:mobius}
\end{figure}

We will now analyse the dynamics of D0-branes in this background, and 
study the \Mth limit in which the model is assumed to be lifted to 
eleven dimensions. As is well known the full dynamics on the compact 
space is described by a gauge theory in $2+1$ dimensions, which is 
obtained by T-dualising the D0-branes \cite{T}. Since the T-duals of 
the shift symmetries are not obvious to us, we find it convenient to 
describe the system slightly differently. Make a change of coordinates to
\be
X_\pm= \fr{X_2}{R_2}\pm \fr{X_1}{R_1}.
\ee 
If we include into the orbifold group the translations that make the 
${\bf R}^2$ into a torus, the group is generated by
$$
\Sh1\Sh2, \Sh1^{-1}\Sh2,\mbox{ and } \Omega\inv1.\
$$
The first two elements shift $X_+$ resp. $X_-$ by one unit, creating 
an ordinary torus out of the $X_+X_-$-plane. The effect of $\inv1$ is 
to exchange the two coordinates:
$$
\inv1\left(\begin{array}{c}X_+\\X_-\end{array}\right) = 
\left(\begin{array}{c}X_-\\X_+\end{array}\right),
$$
and $\Omega$ again exchanges left and right movers. 
In terms of these coordinates we therefore have a type IIA theory on a 
torus, divided out by one (unconventional) symmetry. 
The operation is drawn in figure \ref{fig:mobius}; taking the shaded region as fundamental 
domain, instead of one of the triangles, one easily verifies that this 
indeed represents a M\"obius strip. 
 
The metric on the 
torus is, in terms of the original radii $R_{1,2}$,
\be
\label{eq:metric}
G=\fr14\left(\begin{array}{cc}R_2^2+R_1^2&R_2^2-R_1^2\\
                              R_2^2-R_1^2&R_2^2+R_1^2
             \end{array}\right).
\ee
In particular, when the two radii of the original torus are equal ($R$), 
the new radii are $\tilde{R}=\fr{R}{\sqrt{2}}$. 

The orientifold plane is now really only one 
plane, situated along the diagonal $X_+=X_-$. This diagonal is the 
boundary of the \MS. We again compute the open string spectrum to 
verify that this is indeed the correct model. In the Klein bottle 
calculation, the oscillator contributions are the same as before; 
furthermore we have a momentum sum with $p_+=p_-$, and a winding mode 
sum with $w_+=-w_-$. There are no twisted sectors, since the 
orientifold group has only the one element $\Omega\inv1$. The resulting 
Klein bottle amplitude is
\be
{\cal A}_{KB}=-\frac{8 V_8}{(8\pi^2\alpha')^4}\int dl (16)^2.
\ee
To find the normalisation we also compute the cylinder diagram (for 
convenience we take the case where $R_1=R_2$). Here we have diagonal 
momenta (along the eight-branes) $p=\frac{n}{\sqrt2 \tilde R}$, and 
winding (transverse to the eight-branes) over a multiple of
$\frac{\tilde R}{\sqrt 2}$. The amplitude is
\be
{\cal A}_{Cyl}= -\frac{8 V_8}{(8\pi^2\alpha')^4}\int dl (\Tr 
\gamma_1)^2.
\ee
We clearly need only sixteen D8-branes, distributed symmetrically 
around the orientifold plane. The factor of two difference with the 
usual result comes from the difference between winding and momentum 
sums in the Klein bottle and cylinder. In this representation we therefore obtain a somewhat simpler picture, 
where again the maximal symmetry is $SO(16)$. 

Now we go to the \Mth description. We have to consider the quantum 
mechanics of $N$ D0-branes, with masses $\fr{1}{R_{11}}$,  on the torus 
modded out by the orientifold 
group. In the limit where $N\to\infty$ this will describe M-theory 
on the \MS in the infinite momentum frame. 

To construct the theory we first take the model describing 
zero-branes of type IIA on a torus, and then divide out the symmetry. 
Until further notice we take $R_1=R_2=R$.
The dynamics is described by a $2+1$ $U(N)$ SYM theory with sixteen 
supersymmetries, whose action is the dimensional reduction from 
the ten-dimensional ${\cal N}=1$ gauge theory:
\be
S_{IIA}= -\fr1{4\gym^2}\int \Tr \left( F_{\mu\nu}^2 + 2 \gym^2 D_\mu X_i^2 
-   \gym^4 [X_i,X_j]^2 + \mbox{ fermions} \right).
\ee
The theory contains seven scalars and eight three-dimensional 
fermions, all in the adjoint representation. The two-dimensional space 
is the T-dual of the torus the zero-branes moved on; its radii are
$$
r= \fr{\sqrt{2}\ell_{11}^3}{R_{11}R},
$$
($\ell_{11}$ is the eleven-dimensional Planck length)
and the coupling is 
$$
\gym^2= \fr{2R_{11}}{R^2}.
$$

We now have to factor out the required symmetry $\Omega\inv1$. Under 
T-duality of the $X_+$ and $X_-$ coordinates this symmetry is 
converted into $-\Omega\inv1$. Its fixed planes are therefore the 
lines $X_+=-X_-$, perpendicular to the eight-branes before 
T-dualising, as expected. The action on the fields is as follows:
\bea
\label{eq:transf}
A_0(x_+,x_-) &\to& -A_0^T(-x_-,-x_+) \nn \\
A_+(x_+,x_-) &\to& A_-^T (-x_-,-x_+)  \nn \\
A_-(x_+,x_-) &\to& A_+^T (-x_-,-x_+)  \nn \\
X_i(x_+,x_-) &\to& X_i^T (-x_-,-x_+)  \nn \\ 
\psi_I (x_+,x_-) &\to&  
\fr12\sqrt{2}\gamma_+(1+\gamma_0)\psi_I^T(-x_-,-x_+) .
\eea
The transformation rule of the fermions is motivated by the fact that 
the action of $-\inv1$ can be obtained by first rotating the $X_+X_-$ 
plane over $-\pi/2$ and then reflecting in the line $X_-=0$. Choosing 
three-dimensional gamma-matrices
$$
\gamma_0=i\sigma_2, \quad \gamma_+=\sigma_1,\quad \gamma_-=\sigma_3,
$$
we find the rotation is represented by
$$
\exp{i\fr{\pi}{2}(\fr{i}{2}\gamma_+\gamma_-)}=\left(\begin{array}{cc}  
\fr12\sqrt{2} & \fr12\sqrt{2}\\ -\fr12\sqrt{2}& \fr12\sqrt{2} 
\end{array}\right)=\fr12\sqrt{2}(1+i\sigma_2),
$$    
while the reflection is implemented by multiplication by $\gamma_+$.

These transformations relate the fields on both sides of the fixed 
plane. (Note that these sides are actually connected). In particular 
they impose conditions on the fields living on the fixed plane, 
breaking the $U(N)$ symmetry to a subgroup. Effectively the 
gauge theory itself lives on a M\"obius strip, with specific conditions 
on the fields on the boundary.

Let us then determine the two-dimensional spectrum living on the fixed 
line, obtained by restricting the three-dimensional fields to the 
boundary. First of all we have a two-dimensional vector, consisting of 
the three-dimensional vectors tangent to the line $X_+=-X_-$: $A_0$ 
and $\fr12\sqrt{2}(A_+-A_-)$. From the transformation rules 
(\ref{eq:transf}) we see that these have to be antisymmetric, so that 
the gauge group in two dimensions is broken to $O(N)$. The remaining 
bosonic fields, $\fr12\sqrt{2}(A_++A_-)$ and $X_i$, are in the 
symmetric representation of this group. The three-dimensional fermions 
are split up in two sets of two-dimensional fermions of definite 
chirality. The chirality operator on the fixed line is 
$$
\gamma_3 = \fr12\sqrt{2} \gamma_0(\gamma_+-\gamma_-)= -\fr12\sqrt{2} 
\gamma_+(1+\gamma_0), 
$$ 
so that the spinors of negative chirality are in the symmetric, and 
those of positive chirality in the adjoint representation. 

As in similar models where symmetries with fixed points are divided 
out, we expect extra twisted matter on the fixed line. In the type IA 
description these extra states arise from the quantisation of strings 
connecting the D2-brane to the D8-branes; each two-eight string gives 
rise to one massless fermion, in the fundamental representations of 
the D2 and D8 gauge groups. We expect therefore sixteen Majorana-Weyl 
spinors in the fundamental of $SO(N)$. 

The \Mth motivation for the extra matter is that the two-dimensional 
theory as it stands is anomalous, due to the different representations 
of the left- and right-handed spinors. In fact the anomaly can be 
cancelled by adding $32$ positive chirality fermions in the 
fundamental representation. 

There seems to be a discrepancy between the two ways of counting 
twisted fermions. This can be resolved by making the anomaly argument a 
bit more precise. The gauge anomaly is an anomaly of the 
three-dimensional gauge theory, supported at the boundary. To 
calculate the actual coefficient of the anomaly, following \cite{HW}, 
we perform a gauge transformation whose gauge parameter $\Lambda$ is 
constant along the direction perpendicular to the fixed plane: 
\be
\label{eq:lambda}
(\partial_++\partial_-)\Lambda =0;
\ee
then we will find the two-dimensional anomaly. But now note that    
(\ref{eq:lambda}) implies that $\Lambda$ can be independently defined 
only along half of the fixed line. The line 
$X_++X_-=\mbox{constant}$ intersects the fixed line (or its copies 
under translation) in two different points. We conclude that, by 
symmetry, half of the anomaly is supported on the lower component of the 
boundary, the other half on the upper component. Therefore, to cancel 
the anomaly we indeed only have to add sixteen chiral fermions in the 
fundamental representation.  

We now wish to go to the limit in moduli space where the CHL string is 
weakly coupled, and find the underlying matrix string theory \cite{strings}. 
The coupling constant of the string is 
related to the length of the line segment in the $X_1$ direction,
$$
\lambda_{CHL} = \left( \fr{R_1}{\ell_{11}}\right)^{\fr32}.
$$
In the weak coupling limit we therefore have to send $R_1\to 0$. Let 
us see what this means in terms of the gauge theory. The torus of the 
gauge theory is the T-dual of the $X_+X_-$ torus, which had the metric 
(\ref{eq:metric}). T-dualising means inverting the metric and 
multiplying by $\alpha'^2 = \ell_{11}^6/R_{11}^2$, which yields
\be
\hat{G} = \ell_{11}^6/R_{11}^2 
\left(\begin{array}{cc}R_2^{-2}+R_1^{-2}&R_2^{-2}-R_1^{-2}\\
                              R_2^{-2}-R_1^{-2}&R_2^{-2}+R_1^{-2}
             \end{array}\right).
\ee
So we see that in the weak coupling limit the radii of the torus 
remain equal, and go to infinity, but the angle between the two periodic 
coordinates goes to $-\pi$. 

\begin{figure}
\centerline{\psfig{figure=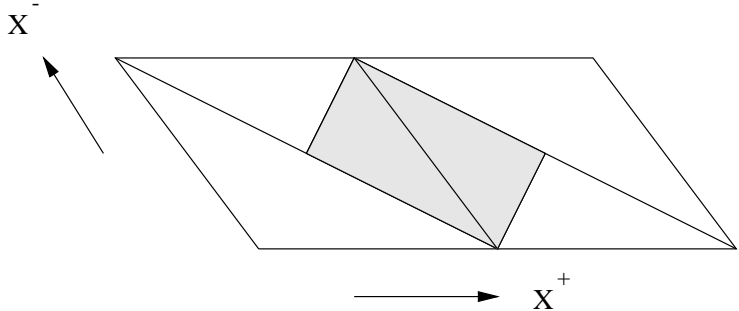}}
\caption{\it In the weak string coupling limit, the strip collapses to a 
line} 
\label{fig:string}
\end{figure}

The fundamental domain (of the torus) therefore degenerates from a 
two-dimensional 
diamond to a one-dimensional line, which will become the string (figure 
\ref{fig:string}). At the same time its surface, $\sqrt{\det \hat{G}}$ 
becomes infinite. 
In this IR limit the (dimensionful) gauge coupling constant also goes to 
infinity.  

The fields living on the line are first of all the untwisted ones. The 
surviving ones are those that are independent of the coordinate 
transverse to the diagonal. We have 
eight scalars and eight right moving fermions, all in the symmetric 
representation of $O(N)$, and furthermore the gauge fields and 
adjoint spinors which are left movers. Then there are the left moving 
twisted fermions in the fundamental of $O(N)$. There are sixteen of 
these, but their periods are twice those of the untwisted fields. 

In the infrared limit the commutator of the $X$-fields is required to 
vanish, so the coordinate fields can be simultaneously 
diagonalised. At the same time the gauge multiplet decouples. The 
eigenvalues of the $X$'s 
may be permuted by a Weyl reflection upon circling the string; 
one so obtains the long strings as twisted sectors. The left moving 
fundamental spinors 
are also exchanged by this Weyl reflection upon completing one cycle 
(which has double the length of a right moving cycle). Furthermore, 
the element $-1$ of $O(N)$ acts non-trivially on the fundamental 
spinors; we will have to project on the subspace of states invariant 
under this element. It is then clear how to obtain the CHL spectrum: 
in the sector with long strings of length $n$, the left moving bosons 
have moding $\fr{1}{n}$. Since the fermions' periods are twice those 
of the bosons, they have half the normal moding: in the anti-periodic 
sector (A) they have modes $\fr{1}{4n}+\fr{m}{2n}$, while in the 
P-sector their modes are $\fr{m}{2n}$. The left moving vacuum energy is 
$-\fr{1}{2n}$ in the A-sector, and $0$ in the P-sector. States of 
zero mass therefore have $SO(16)$ adjoint quantum numbers in the 
A-sector, while in the P-sector they group together in a spinor and an 
anti-spinor of $SO(16)$. One of the latter two is projected out 
by the $-1$ element of $O(N)$, so that we are left with the adjoint 
representation of $E_8$, as expected for the CHL-string.

\newsection{M-theory on a Klein bottle}

For the construction of the matrix model compactified on a Klein bottle 
we start from IIA theory compactified on a two-torus, with radii $R_1$ 
and $R_2$, and then divide out the symmetry $\Omega\inv1\Sh2$ \cite{DP,P}. 
When we lift 
this to eleven dimensions, the operation $\Omega\inv1$ is the inversion 
of the eleventh coordinate. The topology of the M-theory compactification 
manifold is then indeed that of a Klein bottle. In terms of the 
type IIA theory, the inversion of the eleventh dimension is interpreted 
as the operation $(-1)^{F_L}$, with $F_L$ the left moving space-time 
fermion number. This can be checked by comparing the action of both 
operations on the massless fields: they multiply all RR-fields by $-1$.  
In the limit of small eleventh dimension we therefore expect to recover 
a weakly coupled type IIA string, modded out by the symmetry 
$\Sh2(-1)^{F_L}$.

Let us first again review the orientifold model. In the 
closed string sector, the shift $\Sh2$ acts as $(-1)^{n_2}$, with $n_2$ 
the momentum number along the second circle. If we are interested in the 
massless spectrum, we therefore have to find those states invariant under 
$\Omega \inv1$. This is easily done by performing a T-duality along the first 
circle. Then $\Omega\inv1$ is converted to $\Omega$, and we obtain, in 
the massless sector, the closed string spectrum of the type I string. 
There are no closed string twisted sectors, since the only symmetry we 
divide out contains $\Omega$. Then there could be open strings, but upon 
calculating the (world sheet) Klein Bottle contribution to the 
RR-tadpoles, one finds that this vanishes. The reason is that the 
momentum factor in the amplitude is of the form $\sum 
(-1)^ne^{-\fr{t\alpha'n^2}{R_2^2}}$. Poisson resummation shows that this 
vanishes in the $t\to 0$ limit. In conclusion, the theory is consistent 
without D-branes. This is consistent with the observation that the 
symmetry we divided out has no fixed points. 

We now want to introduce zero-branes in this set-up. Their dynamics will 
be described by a certain $2+1$-dimensional gauge theory, which in the limit 
$N\to\infty$ should capture the dynamics of the whole theory. The usual 
strategy to find the appropriate gauge theory is to use T-duality, as we 
did in the previous section. There we could circumvent the problem of 
T-dualising the half shift $\Sh2$ by adopting suitable new coordinates. 
In the present case however, we found no such obvious mechanism to 
T-dualise. 

Instead, we will construct the 
corresponding two-dimensional gauge theory following the original 
strategy of \cite{T,GRT}. In the matrices $X^\mu$ for $N$ zero-branes on a 
compact space we include entries for the images of the zero-branes under 
translation over the circles (and strings wrapping the circles), so that the 
$X^\mu$ are 
infinite-dimensional. The invariance under the translations then poses 
restrictions on the various sub-matrices. In the 
case of zero-branes on a torus, with no further symmetries divided out, 
one finds that in the compact directions the $X^\mu$-matrices have the 
structure 
of a covariant derivative, acting on a field in the fundamental 
representation of $U(N)$. The various $N\times N$ blocks in the infinite 
matrices then correspond to the Fourier components of the gauge fields. 

In the case at hand we also want to divide out the symmetry 
$\Omega\inv1\Sh2$. The shift $\Sh2$ has the effect of adding one to the 
indices denoting the image along the two-direction, plus increasing the 
diagonal entry of $X_2$ by $\pi R_2$. $\Omega$ transposes the  
matrices, while $\inv1$ multiplies $X_1$, and the indices denoting the 
image along the one-direction, by $-1$. 

We demand the matrices to be invariant under this transformation, which 
places restrictions on the various blocks. It turns out that we can again 
represent them as covariant derivatives, 
acting on two fields in 
the fundamental representation, one periodic along the $x_2$ direction 
with radius $\fr{1}{R_2}$, and the other antiperiodic:
$$
\Phi= \bmat{c}\phi^+(n_1,n_2)e^{2\pi i(n_1x_1R_1 + n_2x_2R_2)} \\
                            \phi^-(n_1,n_2)e^{2\pi i(n_1x_1R_1 + 
(n_2+\fr12)x_2R_2)} \emat.
$$
The $X_{1,2}$ can be identified as the Fourier decomposition of the 
covariant derivatives
\be
X_1=-i\partial_1 + \bmat{cc} A_1(x_1,x_2)&B_1(x_1,x_2)\\ 
B^\dag_1(x_1,x_2) & -A_1^T(x_1,-x_2)\emat,\quad
X_2=-i\partial_2 + \bmat{cc} A_2(x_1,x_2)&B_2(x_1,x_2)\\ 
B^\dag_2(x_1,x_2) & A_2^T(x_1,-x_2)\emat,
\ee
with $B_\mu$ {\em antiperiodic} in the $x_2$ 
direction, and satisfying 
$B_\mu (x_1,x_2)=\pm B_\mu^T(x_1,-x_2)$, 
with the $-$ sign for $\mu=1$, and the $+$ for $\mu=2$. The other, 
non-compact, directions have the same structure as $X_2$, but lack of 
course the derivative. 

In this representation, the translations along the torus are represented 
by the unitary transformations 
$$
U_1 = \bmat{cc} e^{2\pi ix_1R_1}&0\\0&e^{2\pi ix_1R_1}\emat,\quad 
U_2=\bmat{cc}e^{2\pi ix_2R_2}&0\\0&e^{2\pi ix_2R_2}\emat,
$$
while the operation $\Sh2$ is given by
$$
U_{\Sh2}=\bmat{cc}0&e^{\pi ix_2R_2}\\
e^{\pi ix_2R_2}&0\emat,
$$
which duly squares to $U_2$.

The fermions behave similarly. We have $16$ fermionic coordinates 
(matrices) $S_\alpha$. The symmetries are the same, except that $\inv1$ 
acts as the ten-dimensional gamma matrix $\gamma_1$ on the ten-dimensional 
(chiral) spinor index. If we then choose a basis in which $\gamma_1$ is 
diagonal, we find that half of the fermions have the structure of $X_1$ 
(the gauginos) and half of them are in the other representation (matter). 

The gauge theory describing M-theory on a Klein Bottle is quite strange. 
Whereas in the case of the M\"obius strip we found a gauge theory whose 
base manifold was again a M\"obius strip, in the Klein Bottle case this 
is definitely not so. Rather, we have gauge fields having some part (the 
$A$-field) living on a torus, without any further restriction beside 
being hermitian. The 
off-diagonal blocks, $B,B^\dag$, the antiperiodic modes of the gauge 
field (or 
the odd modes when going to a circle of double the radius), only have 
independent components on half 
of the torus; effectively they live on a cylinder, and are forced to be 
symmetric on one boundary and anti-symmetric on the other. 
  
We now still want to identify the limit of the theory in which it 
describes a weakly coupled type IIA string. This limit corresponds to 
$R_1\to 0$. In the gauge theory we see that this implies that $x_1$'s 
period goes to infinity. So again, as expected, we find that the 
dimensionless gauge coupling diverges, and that only the zero modes in 
the $x_2$ direction survive. This means that we effectively obtain a 
two-dimensional gauge theory, which we interpret as the world sheet 
theory of a string. Since the $B$ fields are anti-periodic in the 
$x_2$ direction, they do not have zero modes, so their masses go to 
infinity and they can be ignored. The massless fields that remain on 
the world sheet are therefore the regular unitary components $A$. The 
world sheet theory is therefore that of the type IIA string, as we 
expected. 

Then we have to show that this theory indeed is invariant under the 
extra symmetry. First of all it is clearly invariant under the large 
gauge transformation $U_2$, identifying $X_2$ with $X_2+2\pi R_2$, so that 
the theory indeed lives on a circle. To see that the other symmetry 
indeed reduces to $\Sh2(-1)^{F_L}$, note that in the strong coupling 
gauge theory we are considering, the vanishing of the potential 
implies that all the fields can be simultaneously diagonalised. 
Therefore the matter fields are invariant under $U_{\Sh2}$, sending 
$X_2$ to $X_2+\pi R_2$. The gauge field and the gauginos, on the 
other hand, get a $-1$, so to have a full symmetry we need to add to 
$\Sh2$ a transformation multiplying these by an extra $-1$. In the 
strong coupling limit we are considering, the gauge field drops out 
of the action, so basically we have to check that the fermions with 
$\gamma_1$ equal to $-1$ are precisely the fermions of definite 
chirality in the dimensionally reduced gauge theory. 

This can be easily verified by writing the zero-brane lagrangian and 
inserting the solutions for the $X_\mu$ and $S_\alpha$. In a basis 
for the ten-dimensional gamma-matrices with 
$$
\Gamma_0={\bf 1}\otimes i\sigma_2,\quad 
\Gamma_i=\gamma_i\otimes\sigma_1,
$$
with $\gamma_i$ a set of nine-dimensional gamma-matrices,
we have that the ten-dimensional chiral spinors are of the form 
$(S_\alpha,0)$. The interaction term in the matrix model lagrangian 
is then of the form
\be
\Tr \left(-S^T\gamma^i[X_i,S]\right) .
\ee
Inserting the representations we found above, the 
$1+1$-dimensional fermion derivative terms reduce to 
\be
\int dx_1 \Tr \left( -iS^T(\partial_0-\gamma^1\partial_1)S\right),
\ee
so that indeed $\gamma^1$ determines the two-dimensional chirality 
of the spinor. In the weak coupling limit the symmetry therefore can be 
correctly identified as $\Sh2(-1)^{F_L}$.

\newsection{Conclusion}

We have derived the gauge theory models describing matrix theory on 
two non-orientable surfaces, the M\"obius strip and the Klein 
bottle. The M\"obius gauge theory lives on a M\"obius strip itself. In 
the bulk we have a $U(N)$ $2+1$-dimensional gauge theory. On the 
boundary the symmetry is reduced to $O(N)$, with a matter multiplet in the 
symmetric representation. Anomaly cancellation requires extra twisted 
chiral fields on the boundary. In the limit of a small 
line segment the chiral fermions represent the level two $E_8$ current 
algebra living on the CHL string. 

For the Klein bottle compactification, the T-dualisation necessary for 
finding the gauge theory was less straightforward; we had to find a 
covariant derivative representation 
for the zero-brane coordinate matrices ``by hand''. The base manifold of 
the gauge theory turns out to be a hybrid of a torus for the periodic 
modes of the fields and a cylinder for the anti-periodic modes in the 
$x_2$-direction. Alternatively one might describe the theory as living on 
a cylinder, with two gauge fields that are identified on the boundaries, 
and another field that on one boundary is symmetric, on the other 
anti-symmetric. In the limit of weak type IIA string coupling,  
the anti-periodic modes do not survive the dimensional reduction, and we 
are left with a type IIA matrix string invariant under the extra symmetry 
$\Sh2(-1)^{F_L}$.   

\acknowledgements

The author is indebted to Erik Verlinde for advice and discussions, 
and to FOM for financial support.

\end{document}